\documentclass[12pt,twoside]{article}

\begin{document}

\title{One Hundred Years of Quantum Physics}
\author{Daniel Kleppner and Roman Jackiw\\
Massachusetts Institute of Technology\\
\small MIT-CTP\#3016\\
\small (To be published in \emph{Science}, August 11, 2000)
}
\date{}
\maketitle
\thispagestyle{empty}

\begin{abstract}\noindent
 On the occasion of the 100th anniversary of the birth of the quantum
 idea, the development, achievements, and promises of quantum mechanics
 are described.
\end{abstract}

An informed list of the most profound scientific developments 
in the twentieth century is likely to include general relativity, 
quantum mechanics, big-bang cosmology, the unraveling of the 
genetic code, evolutionary biology, and perhaps a few other topics of 
the reader's choice. Among these quantum mechanics is unique 
because of its profoundly radical quality. Quantum mechanics forced 
physicists to reshape their ideas of reality, to rethink the nature of 
things at the deepest level, to revise their concepts of position and 
speed, their notions of cause and effect.

Although quantum mechanics was created to describe an 
abstract atomic world far removed from daily experience, its impact 
on our daily lives could hardly be greater. The spectacular advances 
in chemistry, biology, and med\-i\-cine -- and in essentially every other 
science -- could not have occurred without the tools that quantum 
mechanics made possible. Without quantum mechanics there would 
be no global economy, for the electronics revolution that brought us 
the computer age is a child of quantum mechanics, as is the 
photonics revolution that brought us the information age. The 
creation of quantum physics has transformed our world, bringing 
with it all the benefits -- and the risks -- of a scientific revolution.  

Unlike general relativity, which grew out of a brilliant insight 
into the connection between gravity and geometry, or the 
deciphering of DNA, which unveiled a new world of biology, quantum 
mechanics did not spring from a single step. Rather, it was created 
by a small group of physicists in one of those rare concentrations of 
genius that occur from time to time in history. Following a period of 
twenty years when quantum ideas had been introduced but were so 
confused that there was little basis for progress, they created 
quantum mechanics in three tumultuous years. They were troubled by 
what they were doing and in some cases distressed by what they had 
done. 

The unique situation of this crucial yet elusive theory is 
perhaps best summarized by the following observation: Quantum 
theory is the most precisely tested and most successful theory in 
the history of science.  Nevertheless, not only was quantum 
mechanics deeply disturbing to its founders, today -- seventy-five 
years after the theory was essentially cast in its current 
form -- some of the luminaries of science remain dissatisfied with 
its foundations and its interpretation, even as they acknowledge its 
stunning power.

This year marks the hundredth anniversary of Max Planck's 
creation of the quantum concept. In his seminal paper on thermal 
radiation, Planck hypothesized that the total energy of a vibrating 
system cannot be changed continuously. Instead, the energy must 
jump from one value to another in discrete steps, or quanta, of 
energy. The idea of energy quanta was so radical that Planck let it 
lay fallow. Then Einstein, in his wonder year of 1905, recognized 
the implications of quantization for light. Even then the concept was 
so bizarre that there was little basis for progress.  Twenty more 
years and a fresh generation of physicists were required to create 
modern quantum theory. 

To understand the revolutionary impact of quantum physics one 
need only look at pre-quantum physics. In the years 1890--1900 the 
journals of physics were filled with papers on atomic spectra and 
essentially every other measurable property of matter such as 
viscosity, elasticity, electrical and thermal conductivity, 
coefficients of expansion, indices of refraction, and thermoelastic 
coefficients. Spurred by the energy of the Victorian work ethic and 
the development of ever more ingenious experimental methods, 
knowledge accumulated at a prodigious rate. What is most striking 
to the contemporary eye, however, is that the compendious  
descriptions  of the properties of matter were essentially empirical. 
Thousands of pages of spectral data listed precise values for the 
wavelengths of the elements, but nobody knew why spectral lines 
occurred, much less what information they conveyed. Thermal and 
electrical conductivities were interpreted by suggestive models 
that fitted roughly half of the facts. There were numerous empirical 
laws but they were not satisfying. For instance, the law of Dulong 
and Petit established a simple relation between specific heat and 
atomic weight of a material. Much of the time it worked; some of the 
time it did not. The masses of equal volumes of gas were in the 
ratios of integers -- mostly. The Periodic Table, which provided a key 
organizing principle for the flourishing science of chemistry, had absolutely
no theoretical basis. 

Among the greatest achievements of the revolution to come is  
this: Quantum mechanics provides a quantitative theory of matter. 
We now understand essentially every detail of atomic structure; the 
Periodic Table has a simple and natural explanation; the vast arrays 
of spectral data fit into an elegant theoretical framework. Quantum 
theory permits the quantitative understanding of molecules, solids 
and liquids, of conductors and semiconductors. It explains bizarre 
phenomena such as superconductivity and superfluidity, exotic forms 
of matter such as the stuff of neutron stars and Bose-Einstein 
condensates in which all the atoms in a gas behave like a single 
super atom. Quantum mechanics provides essential tools for all of 
the sciences and for every advanced technology. 

Quantum physics actually encompasses two entities.  The first 
is the theory of matter at the atomic level -- quantum mechanics. It is 
quantum mechanics that allows us to understand and manipulate the 
material world. The second is the quantum theory of fields. Quantum 
field theory plays a totally different role in science, to which we 
shall return.

\subsection*{Quantum Mechanics}

The clue that triggered the quantum revolution came not from 
the studies of matter, but from a problem in radiation. The specific 
challenge was to understand the spectrum of light emitted by hot 
bodies -- blackbody radiation. The phenomenon is familiar to anyone 
who has stared at a fire. Hot matter glows and the hotter it becomes 
the brighter it glows. The spectrum of the light is broad with a peak 
that shifts from red to yellow and finally to blue (though we cannot 
see that) as the temperature is raised. It should have been possible 
to understand the shape of the spectrum by combining concepts from 
thermodynamics and electromagnetic theory, but all attempts failed. 
However, by assuming that the energies of the vibrating electrons 
that radiate the light are quantized, Planck obtained an expression 
that agreed beautifully with experiment. But as he recognized all too 
well, the theory was physically absurd, ``an act of desperation'' as he 
later described it.   

Planck applied his quantum hypothesis to the energy of the 
vibrators in the walls of a radiating body. Quantum physics might 
have ended there if in 1905 a novice -- Albert Einstein -- had not 
reluctantly concluded that if a vibrator's energy is quantized then 
the energy of the electromagnetic field that it radiates -- light -- must 
also be quantized. Einstein thus imbued light with particle-like 
behavior, notwithstanding that Maxwell's theory, and over a century 
of definitive experiments, testified to light's wave nature. 
Experiments on the photoelectric effect in the following decade 
revealed that when light is absorbed its energy actually arrives in 
discrete bundles, as if carried by a particle. The dual nature of 
light -- particle-like or wave-like depending on what one looks 
for -- was the first example of a vexing theme that would recur 
throughout quantum physics. The duality constituted a theoretical 
conundrum for the next twenty years. 

The first step towards quantum theory had been precipitated 
by a dilemma about radiation. The second step was precipitated by a 
dilemma about matter. It was known that atoms contain positively 
and negatively charged particles. But oppositely charged particles 
attract. According to electromagnetic theory, therefore, they should 
spiral into each other, radiating light in a broad spectrum until they 
collapse. Once again, the door to progress was opened by a 
novice -- Niels Bohr.  In 1913 Bohr proposed the radical hypothesis 
that electrons in an atom exist only in certain stationary states, 
including a ground state.  Electrons change their energy by ``jumping'' 
between the stationary states, emitting light whose wavelength 
depends on the energy difference.  By combining known laws with 
bizarre assumptions about quantum behavior, Bohr swept away the 
problem of atomic stability. Bohr's theory was full of contradictions 
but it provided a quantitative description of the spectrum of the 
hydrogen atom. He recognized both the success and the shortcomings 
of his model. With uncanny foresight, Bohr rallied physicists to 
create a new physics. His vision was eventually fulfilled, though it 
took twelve years and a new generation of young physicists.  

At first, attempts to advance Bohr's quantum ideas -- the so-called old
quantum theory -- suffered one defeat after another. Then a  series of
developments totally changed the course of thinking. 

In 1923 Louis deBroglie, in his PhD thesis, proposed that the 
particle behavior of light should have its counterpart in the wave 
behavior of particles. He associated a wavelength with the 
momentum of a particle -- the higher the momentum the shorter the 
wavelength. The idea was intriguing, but no one knew what a 
particle's wave nature might signify or how it related to atomic 
structure. Nevertheless, deBroglie's hypothesis was an important 
precursor for events soon to come. 

In the summer of 1924 there was yet another precursor. 
Setyendra N. Bose proposed a totally new way to explain the Planck 
radiation law.  He treated light as if it were a gas of massless 
particles (now called photons) that do not obey the classical laws 
of Boltzmann statistics but a type of new statistics based on their 
indistinguishable nature. Einstein immediately applied Bose's 
reasoning to a real gas of massive particles and obtained a new law 
 -- to become known as the Bose-Einstein distribution -- for  how 
energy is shared  by the particles in a gas. However, under normal 
circumstances the new and old theories predicted the same behavior 
for atoms in a gas. Einstein took no further interest and the result 
lay undeveloped  for more than a decade. Its key idea, however, the 
indistinguishability of particles, was about to become critically 
important.

Suddenly, a tumultuous series of events occurred that 
culminated in a scientific revolution. In the three-year period from 
January 1925 to January 1928:

\begin{itemize}

\item  Wolfgang Pauli proposed the exclusion principle, providing a 
theoretical basis for the Periodic Table.
  
\item  Werner Heisenberg, with Max Born and Pascual Jordan, 
discovered matrix mechanics, the first version of quantum 
mechanics. The historical goal of understanding electron motion 
within atoms was abandoned in favor of a systematic method for 
organizing observable spectral lines. 
 
\item  Erwin Schr\"odinger invented wave mechanics, a second form of 
quantum mechanics in which the state of a system is described 
by a wave function, the solution to Schr\"odinger's equation. 
Matrix mechanics and wave mechanics, apparently incompatible, 
were shown to be equivalent.
 
\item  Electrons were shown to obey a new type of statistical law, 
Fermi-Dirac statistics. It was recognized that all particles obey 
either Fermi-Dirac statistics or Bose-Einstein statistics, and 
that the two classes have fundamentally different properties.
 
\item  Heisenberg enunciated the Uncertainty Principle. 
 
\item  Paul A.M. Dirac developed a relativistic wave equation for the 
electron that explained electron spin and predicted anti-matter.
 
\item  Dirac laid the foundations of quantum field theory by providing 
a quantum description of the electromagnetic field.
 
\item  Bohr announced the complementary principle, a philosophical 
principle that helped to resolve apparent paradoxes of quantum 
theory, particularly the wave-particle duality.

\end{itemize}

The principal players in the creation of quantum theory were 
young. In 1925, Pauli was 25 years old, Heisenberg 24, Dirac 23, 
Jordan 23, Fermi 24. Schr\"odinger, at 36 years, was a late bloomer. 
Born and Bohr were older yet and it is significant that their 
contributions were largely interpretative. The profoundly radical 
nature of the intellectual achievement is revealed by Einstein's 
reaction. Having invented some of the key concepts that led to 
quantum theory, Einstein rejected it. His paper on Bose-Einstein 
statistics was his last contribution to quantum physics and his last 
significant contribution to physics. 

That a new generation of physicists was needed to create 
quantum mechanics is hardly surprising. Lord Kelvin described why 
in a letter to Bohr congratulating him on his 1913 paper on hydrogen. He said
that there  was  much truth in Bohr's paper, but he would  never understand
it himself. Kelvin recognized that radically new  physics would need to come
from unfettered minds. 

In 1928 the revolution was finished and the foundations of 
quantum mechanics were essentially complete. The frenetic pace 
with which it occurred  is revealed by an anecdote recounted by 
Abraham  Pais\footnote{\emph{Inward Bound}, Oxford University Press,
1986}.   In 1925 the concept of electron spin had been proposed by Samuel 
Goudsmit and George Uhlenbeck. Bohr was deeply skeptical. In 
December he traveled to Leiden to attend the jubilee of Hendrik A. 
Lorentz's doctorate. Pauli met the train at Hamburg to find out 
Bohr's opinion about the possibility of electron spin. Bohr said the 
proposal was ``very, very interesting,'' his well-known put down 
phrase. Later at Leiden, Einstein and Paul Ehrenfest met Bohr's train, 
also  to discuss spin. There, Bohr explained his objection, but 
Einstein showed a way around it and converted Bohr into a supporter. 
On his return journey, Bohr met up with yet more discussants. When 
the train passed through G\"ottingen, Heisenberg and Jordan were 
waiting at the station to ask his opinion. And at the Berlin station, 
Pauli was waiting, having traveled specially from Hamburg. Bohr 
told them all that the discovery of the electron spin was a great 
advance. 

The creation of quantum mechanics triggered a scientific gold 
rush. Among the early achievements were these: Heisenberg laid the 
foundations for atomic structure theory by obtaining an approximate 
solution to Schr\"odinger's equation for the helium atom in 1927, and 
general techniques for calculating the structures of atoms were 
created soon after by John Slater, Douglas Rayner Hartree, and 
Vladimir Fock. The structure of the hydrogen molecule was solved by 
Fritz London and Walter Heitler; Linus Pauling built on their results 
to found theoretical chemistry. Arnold Sommerfeld and Pauli laid the 
foundations of the theory of electrons in metals and Felix Bloch 
created band-structure theory. Heisenberg explained the origin of 
ferromagnetism. The enigma of the random nature of radioactive 
decay by alpha-particle emission was explained in 1928 by George 
Gamow, who showed that it occurs by quantum mechanical tunneling. In the
following years Hans Bethe laid the foundations for  nuclear physics and
explained the energy source of stars. With these  developments atomic,
molecular, solid state, and nuclear physics  entered the modern age.

\subsection*{Controversy and Confusion}

Side by side with these advances, however,  fierce debates 
were taking place on the interpretation and validity of quantum 
mechanics. Foremost among the protagonists were Bohr and 
Heisenberg, who embraced the new theory, and Einstein and 
Schr\"odinger, who were dissatisfied. To appreciate the reasons for 
such turmoil one needs to understand some of the key features of 
quantum theory, which we summarize here. For simplicity, we 
describe the Schr\"odinger version of quantum mechanics, sometimes 
called wave mechanics.

\paragraph{Fundamental description: The wave function} The behavior of a 
system is described by Schr\"odinger's equation. The solutions to 
Schr\"odinger's equation are known as wave functions. The complete 
knowledge of a system is described by its wave function and from 
the wave function one can calculate the possible values of every 
observable quantity. The probability of finding an electron in a given 
volume of space is proportional to the square of the magnitude of 
the wave function. Consequently, the location of the particle is 
``spread out'' over the volume of the wave function. The momentum of 
a particle depends on the slope of the wave function; the greater the 
slope, the higher the momentum. Because the slope varies from place 
to place, momentum is also ``spread out.'' The need to abandon a 
classical picture in which position and velocity can be determined 
with arbitrary accuracy in favor of a blurred picture of probabilities 
is at the heart of quantum mechanics. 

Measurements made on identical systems that are identically 
prepared will not yield identical results. Rather, the results will be 
scattered over a range described by the wave function. Consequently, 
the concept of an electron having a particular location and a 
particular momentum looses its foundation. The uncertainty 
principle quantifies this: To locate a particle precisely the wave 
function must be sharply peaked (that is, not spread out). However a 
sharp peak requires a steep slope, and so the spread in momentum 
will be great. Conversely, if the momentum has a small spread, the 
slope of the wave function must be small, which means that it must 
spread out over a large volume.

\paragraph{Waves can interfere}Their heights add or subtract depending 
on their relative phase. Where the amplitudes are in phase, they add; 
where they are out of phase, they subtract. If a wave can follow 
several paths from source to receiver, as a light wave undergoing 
two-slit interference, then the illumination will generally display 
interference fringes. Particles, obeying a wave equation will do 
likewise, as in electron diffraction. The analogy seems reasonable 
until one inquires about the nature of the wave. A wave is generally 
thought of as a disturbance in a medium. In quantum mechanics there 
is no medium, and in a sense there is no wave since the wave 
function is fundamentally a statement of our knowledge of a system. 

\paragraph{Symmetry and identity} A helium atom consists of a nucleus 
surrounded by two electrons. The wave function of helium describes 
the position of each electron. However, there is no way of 
distinguishing which electron is which. Consequently, if the 
electrons are switched the system must look the same, which is to 
say the probability of finding the electrons in given positions is 
unchanged. Because the probability depends on the square of the 
magnitude of the wave function, the wave function for the system 
with the interchanged particles must be related to the original wave 
function in one of two ways: Either it is identical to the original 
wave function, or it's sign is simply reversed, i.e.,  it is multiplied 
by a factor $-1$.  Which one is it?

One of the astonishing discoveries in quantum mechanics is 
that for electrons the wave function always changes sign. The 
consequences are dramatic, for if two electrons were in the same 
quantum state, then the wave function would have to be its negative 
opposite. Consequently, the wave function must vanish. Thus, the 
probability of finding two electrons in the same state is zero. This 
is the Pauli exclusion principle.  All particles with half-integral 
spin, including electrons, behave this way and are called fermions. 
For particles with integer spin, including photons, the wave function 
does not change sign. Such particles are called bosons. Electrons in 
an atom arrange themselves in shells because they are fermions but 
light from a laser emerges in a single super intense 
beam -- essentially a single quantum state -- because light is composed 
of bosons. Recently, atoms in a gas have been cooled to the quantum 
regime where they form a Bose-Einstein condensate in which the 
system can emit a super intense matter beam -- forming an atom 
laser. 

These ideas apply only to identical particles since if different 
particles are interchanged the wave function will certainly be 
different. Consequently, particles behave like fermions or like 
bosons only if they are totally identical. The absolute identity of 
like particles is among the most mysterious aspects of quantum 
mechanics. Among the achievements of quantum field theory is that 
it can explain this mystery. 

\paragraph{What does it mean?} Questions such as what  a wave function 
``really is'' and what is meant by ``making a measurement'' were 
intensely debated in the early years.  By 1930,  however,  a more or 
less standard interpretation of quantum mechanics had been 
developed by Bohr and his colleagues, the so-called Copenhagen 
interpretation. The key elements are the probabilistic description of 
matter and events, and reconciliation of the wave-like and particle-like
natures of things through Bohr's principle of complementarity.  Einstein never
accepted quantum theory. He and Bohr debated its  principles until Einstein's
death in 1955. 

A central issue in the debates on quantum mechanics was 
whether the wave function contains all possible information about a 
system or if there might be underlying factors -- hidden 
variables -- that determine the outcome of a particular measurement. 
In the mid-1960s John S. Bell showed that if hidden variables 
existed, experimentally observed probabilities would have to fall 
below certain limits, dubbed ``Bell's inequalities.'' Experiments were 
carried out by a number of groups, which found that the inequalities 
were violated. Their collective data came down decisively against 
the possibility of hidden variables. For most scientists this resolved 
any doubt about the validity of quantum mechanics. 

	Nevertheless, the nature of quantum theory continues to 
attract attention because of the fascination with what is sometimes 
described as ``quantum weirdness.'' The weird properties of quantum 
systems arise from what is known as entanglement. Briefly, a 
quantum system, such as an atom, can exist in any one of a number 
of stationary states but also in a superposition -- or sum -- of such 
states. If one measures some property such as the energy of an atom 
in a superposition state, in general the result is sometimes one 
value, sometimes another. So far, nothing is weird.  

It is also possible, however, to construct a two-atom system 
in an entangled state in which the properties of both atoms are 
shared with each other. If the atoms are separated, information 
about one is shared, or entangled, in the state of the other. The 
behavior is unexplainable except in the language of quantum 
mechanics. The effects are so surprising that they are the focus of 
study by a small but active theoretical and experimental community. 
The issues are not limited to questions of principle, since 
entanglement can be useful. Entangled states already have been 
employed in quantum communication systems, and entanglement 
underlies all proposals for quantum computation.

\subsection*{The Second Revolution}

During the frenetic years in the mid-1920s when quantum 
mechanics was being invented, another revolution was underway. The 
foundations were being laid for the second branch of quantum physics --
quantum field theory. Unlike quantum mechanics, which was  created in a
short flurry of activity and emerged essentially  complete, quantum field
theory has a tortuous history that continues  today. In spite of the difficulties,
however, the predictions of  quantum field theory are the most precise in all
of physics, and  quantum field theory constitutes a paradigm for some of the
most  crucial areas of theoretical inquiry. 

The problem that motivated quantum field theory was the 
question of how an atom radiates light as its electrons ``jump'' from 
an excited states to its ground state. Einstein proposed such a 
process, called spontaneous emission, in 1916, but he had no way to 
calculate its rate. Solving the problem required developing a fully 
relativistic quantum theory of electromagnetic fields, a quantum 
theory of light. Quantum mechanics is the theory of matter. Quantum 
field theory, as its name suggests, is the theory of fields, not only 
electromagnetic fields but other fields that were subsequently discovered. 

In 1925 Born, Heisenberg, and Jordan published some initial 
ideas for a theory of light, but the seminal steps were due to 
Dirac -- a young and essentially unknown physicist working in 
isolation -- who presented his field theory 1926. The theory was full 
of pitfalls: formidable calculational complexity, predictions of 
infinite quantities, and apparent violations of the correspondence 
principle. In the late 1940s a new approach to the quantum theory of 
fields, QED (for quantum electrodynamics) was developed by Richard 
Feynman, Julian Schwinger, and Sin-itiro Tomonaga. They 
sidestepped the infinities by a procedure, called renormalization, 
which essentially subtracts infinite quantities so as to leave finite 
results. Because there is no exact solution to the complicated 
equations of the theory, an approximate answer is presented as a 
series of terms that become more and more difficult to calculate. 
Although the terms become successively smaller, at some point they 
should start to grow, indicating the breakdown of the approximation.  
In spite of these perils, QED ranks among the most brilliant 
successes in the history of physics. Its prediction of the interaction 
strength between an electron and a magnetic field has been 
experimentally confirmed to a precision of two parts in 
1,000,000,000,000. 

Notwithstanding its fantastic successes, QED harbors enigmas. 
The view of empty space -- the vacuum -- that the theory provides 
initially seems preposterous. It turns out that empty space is not 
really empty. Rather, space is filled with small fluctuating 
electromagnetic fields. These vacuum fluctuations are essential for 
explaining spontaneous emission. Furthermore, they produce small 
but measurable shifts in the energies of atoms and certain 
properties of particles such as the electron. Strange as they seem, 
these effects have been confirmed by some of the most precise 
experiments ever carried out.

At the low energies of the world around us, quantum mechanics 
is fantastically accurate.  But at high energies where relativistic 
effects come into play, a more general approach is needed.  Quantum 
field theory was invented to reconcile quantum mechanics with 
special relativity. 

The towering role that quantum field theory plays in physics 
arises from the answers it provides to some of the most profound 
questions about the nature of matter. Quantum field theory explains 
why there are two fundamental classes of particles -- fermions and 
bosons -- and how their properties are related to their intrinsic spin. 
It describes  how particles are created and annihilated, not only 
photons, but electrons and positrons (antielectrons). It explains the 
mysterious nature of identity in quantum mechanics -- how identical 
particles are absolutely identical because they are created by the 
same underlying field. QED describes not only the electron but the 
class of particles called leptons that includes the muon, the tau 
meson, and their antiparticles. Because QED is a theory for leptons 
it cannot describe more complex particles called hadrons. These  
include protons, neutrons, and a  wealth of mesons.  For hadrons,  a 
new theory had to be invented, a generalization of QED called 
quantum chromodynamics, or QCD.  Analogies abound between QED 
and QCD. Electrons are the constituents of atoms; quarks are the 
constituents of hadrons.  In QED the force between charged particles 
is mediated by the photon; in QCD the force between quarks is 
mediated by the gluon. In spite of the parallels, however, there is a 
crucial difference between QED and QCD. Unlike leptons and photons, 
quarks and gluons are forever confined within the hadron. They 
cannot be liberated and studied in isolation.

QED and QCD are the cornerstones for a grand synthesis known 
as the standard model. The standard model has successfully 
accounted for every particle experiment carried out to date. 
However, for many physicists the standard model is inadequate 
because data on the masses, charges and other properties of the 
fundamental particles need to be found from experiments. An ideal 
theory would predict all of these.

Today, the quest to understand the ultimate nature of matter 
is at the focus of an intense scientific study that is reminiscent of 
the frenzied and miraculous days in which quantum mechanics was 
created, and whose outcome may be even more far reaching. The 
effort is inextricably bound to the quest for a quantum description 
of gravity. The procedure for quantizing the electromagnetic field 
that worked so brilliantly in QED has failed to work for gravity, in 
spite of a half century of effort. The problem is critical, for if 
general relativity and quantum mechanics are both correct, then they 
must ultimately provide a consistent description for the same 
events. There is no contradiction in the normal world around us because
gravity is so fantastically weak compared to the electrical  forces in atoms
that quantum effects are negligible and a classical  description works
beautifully. But for a system such as a black hole  where gravity is incredibly
strong we have no reliable way to  predict quantum behavior. 

One century ago our understanding of the physical world was 
empirical. Quantum physics gave us a theory of matter and fields, 
and that knowledge transformed our world. Looking to the next 
century, quantum mechanics will continue to provide fundamental 
concepts and essential tools for all of the sciences. We can make 
such a prediction confidently because for the world around us 
quantum physics provides an exact and complete theory. However, 
physics today has this in common with physics in 1900: Physics 
remains ultimately empirical -- we cannot fully predict the properties 
of the elementary constituents of matter, we must measure them. 
Perhaps string theory -- a generalization of quantum field theory that 
eliminates all infinities by replacing point-like objects such as the 
electron with extended objects -- or some theory only now being conceived,
will solve the riddle. Whatever the outcome, the dream for  ultimate
understanding will continue to be a driving force for new  knowledge, as it
has been since the dawn of science. One century  from now, the consequences
of pursuing that dream will belie our  imagination.  

\end{document}